\documentclass{llncs}
\usepackage{amsmath,graphicx}
\usepackage[normalem]{ulem}
\usepackage{url}
\usepackage{mathtools, cuted}
\usepackage[table,x11names,dvipsnames]{xcolor}
\usepackage{caption}
\usepackage{graphics}
\usepackage{tikz}
\usepackage[hidelinks]{hyperref}

\usepackage{url}

\begin{document}
\mainmatter              
\title{Transfer Learning with Jukebox for Music Source Separation}
\titlerunning{Transfer Learning with Jukebox for Music Source Separation}  
%
\author{Wadhah Zai El Amri\and Oliver Tautz\and Helge Ritter \and Andrew Melnik}
\authorrunning{Zai El Amri et al.} 
%
%
\institute{Bielefeld University, Germany\\
\email{wadhah.zai.papers@gmail.com, oliver.tautz.papers@gmail.com, helge@techfak.uni-bielefeld.de, andrew.melnik.papers@gmail.com}
}

\maketitle


\begin{abstract}
In this work, we demonstrate how a publicly available, pre-trained \textit{Jukebox} model can be adapted for the problem of audio source separation from a single mixed audio channel. Our neural network architecture, which is using transfer learning, is quick to train and the results demonstrate performance comparable to other state-of-the-art approaches that require a lot more compute resources, training data, and time. We provide an open-source code implementation of our architecture (\url{https://github.com/wzaielamri/unmix})

\end{abstract}



\section{Introduction and Related Work}

Source separation is an important issue in many fields such as audio processing, image processing \cite{melnik2021critic}, EEG \cite{melnik2017systems,melnik2017eeg}, etc. Music source separation from mixed audio is a challenging problem, especially if the source itself should be learned from a dataset of examples. Additionally, such models are expensive to train from scratch. We tested our model on the MUSDB18-HQ \cite{MUSDB18HQ} dataset which supplies full songs with ground truth stems of 'bass', 'drums', 'vocals' and 'other', which includes instruments such as guitars, synths, etc. The task is to separate a mixed audio channel into the separately recorded instruments, called stems here. Most baseline models in the Music Demixing Challenge 2021 \cite{musicDemixing} used masking of input transformed to the frequency domain by short-time Fourier transformation such as UMX~\cite{Stoeter2019}. This older technique transforms the waveform into a three dimensional input that can be viewed as a picture. The model then computes a 'mask' which substracts parts of the audio in frequency domain before transforming back to listenable audio. \textit{Demucs} \cite{DBLP:journals/corr/abs-1909-01174} on the other hand showed a successful approach that works in waveform domain, where an autoencoder, based on a bidirectional long short-term memory network, is used. \newline
This success of such a solution (using waveforms) encouraged us to adapt \textit{Jukebox} \cite{dhariwal2020jukebox}, a powerful, generative model using multiple, deep Vector Quantized-Variational Autoencoders (VQ-VAE) \cite{DBLP:journals/corr/abs-1711-00937} to automatically generate real sounding music, and using its publicly available pre-trained weights for the task.


Transfer learning helped deep learning models reach new heights in many domains, such as natural language processing \cite{DBLP:journals/corr/abs-1810-04805,DBLP:journals/corr/abs-1910-10683} and computer vision \cite{HAN201843,https://doi.org/10.1111/mice.12363}. Although relatively unexplored for the audio domain, \cite{7472128} proved feature representation learned on speech data could be used to classify sound events. Their results verify that cross-acoustic transfer learning performs significantly better than a baseline trained from scratch. TRILL \cite{Shor_2020} showed great results of pre-training deep learning models with an unsupervised task on a big dataset of speech samples. Its learned representations exceeded SOTA performance on several downstream tasks with datasets of limited size.

We take a similar approach that is heavily based on \textit{Jukebox} \cite{dhariwal2020jukebox}. It uses multiple VQ-VAEs to compress raw audio into discrete codes. They are trained self-supervised, on a large dataset of about 1.2 million songs, needing the compute power of 256 V100 to train in an acceptable time. Our experiments show that \textit{Jukebox's} learned representations can be used for the task of source separation.

\begin{figure*}[ht]
 \centering
 \includegraphics[width=0.9\textwidth]{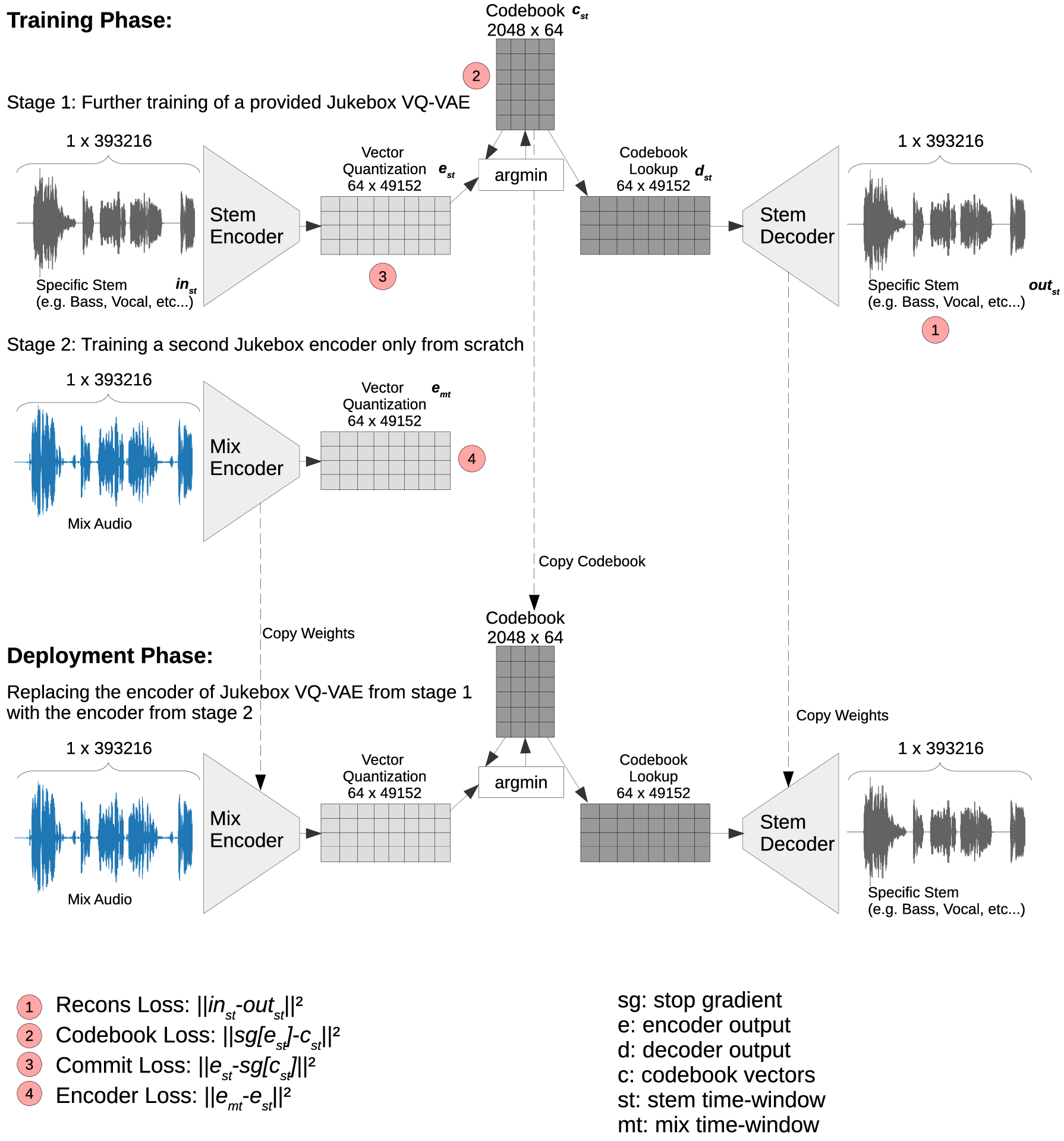}
 \caption{Visualization of the proposed transfer learning model architecture.}
 \label{fig:fig1}
\end{figure*}

\section{Method}
\subsection{Architecture}

Our architecture utilizes the default \textit{Jukebox's} \cite{dhariwal2020jukebox} variant of the publicly available pre-trained VQ-VAE model. \textit{Jukebox} uses three separated VQ-VAEs. We use only the smallest one with the strongest compression, due to the low resource usage. It employs dilated 1-D convolutions in multiple residual blocks to find a less complex sequence representation of music. An audio sequence $x_t$ gets mapped by an encoder $E_1$ to a latent space $e_t=E_1(x_t)$ of 64 dimensions so that it can be mapped to the closest prototype vector in a collection $C$ of vectors called \textit{codebook}. These 2048 prototype vectors, denoted $c_{st}$, are learned in training and help to form a high-quality representation.

The rate of compression for a sequence is called the hop length, for which a value of 8 is used. It depends on the stride values of the convolutional layers. We set the stride value to 2 as well as the downsampling to 3. All other values remain as defined in \cite{dhariwal2020jukebox}. After mapping to the codebook, a decoder $D$ aims to reconstruct the original sequence. In summary, equation (\ref{eq:1})

\begin{equation}
\label{eq:1}
  y_t=D(argmin_{c}(\|E_1(x_t)- c)\|) \;\; \text{for} \;\; c \in C
\end{equation} 

describes a full forward pass through the VQ-VAE, where $ y_t $ is the prediction for an input sequence $x_t$ and $\|.\|$ is the euclidean norm. For further technical details on the used VQ-VAE architecture, refer to the paper of Dhariwal et al\cite{dhariwal2020jukebox}. The model is fine-tuned on data for one stem, learning good representations for a single instrument. In addition, we train a second encoder $E_2$, identical to the one already mentioned, to project an input sequence of the mixture to the space already known by the codebook and decoder. For deployment, the encoder of the VQ-VAE is switched with the second one, effectively mapping from the full mixture to one stem.

\subsection{Data}

Our models are trained on the MUSDB18-HQ \cite{MUSDB18HQ} dataset, also used in the music demixing challenge \cite{musicDemixing}. It consists of 150 full-length songs, sampled at 44KHz, providing the full audio mixture and four stems, 'vocals', 'bass', 'drums', and 'other' for each sample, which can be regarded as ground truth in the context of source separation. We train on the full train set composed of 100 songs, testing is done on the remaining 50. 

\subsection{Training}



For each stem i=1..4, we train a model in two phases (see Fig.~\ref{fig:fig1}). In
the first phase, the model is trained on data that present the chosen
stem in isolation (i.e. not embedded in a mixture). This produces a
VQ-VAE with a "single stem encoder" ($\text{SE}_i$) that can map a single
stem into a good latent representation, followed by a "stem decoder"
($\text{SD}_i$)  tuned  to  reconstruct  the  input  after  the  "discretization
bottleneck" as faithfully as possible. Training of each VQ-VAE is based
on the same three losses as chosen in the original Jukebox paper
\cite{dhariwal2020jukebox} : $L = L_{recons} + L_{codebook} + \beta L_{commit}$. However, our
final goal is to process each stem when it is part of a mixture of all
four stems. Such embedding will introduce distortion of each stem,
requiring to replace each single stem encoder $\text{SE}_i$ from phase 1 by a
corresponding "mixture stem encoder" ($\text{ME}_i$) that is trained in phase
2, to map its changed (mixture embedded) stem $
i$ input onto the representation (stem-i codebook prototypes) created
in phase by the $\text{SE}_{i}$-$\text{SD}_i$ encoder-decoder pair. So, for each stem $i$
(now omitting index $i$ in the following) for each training sample ($x_{mt}$:
the sequence of the mixed audio, $ x_{st}$ : the sequence of stem audio),
we feed  $x_{st}$, to  the  already  trained  encoder SE,  producing  $e_{st}$.
Separately, the full mixture $x_{mt}$ is passed through the new encoder
ME, yielding $e_{mt}$. Now, we can backpropagate through ME using MSE
loss $||e_{st}-e_{mt}||^2$ (keeping SE fixed throughout phase 2). 
It would also be possible to train E2 end-to-end, similar to training phase 1. This could be done in two ways, freezing the weights of the rest of the VQ-VAE or fine-tuning further. With frozen weights, the best reachable performance would be identical to our current approach, but training time
would be increased. Continuing to fine-tune would most probably lead to more source bleeding because the embeddings no longer represent only the stem they were trained on in training phase 1, so we decided against training end-to-end.

Finally, we
obtain  our  mixture-adapted  final  VQ-VAE  by  concatenating  the
trained mixture stem encoder ME with the stem decoder SD. Note
that this procedure will be carried out for each of the four stems,
yielding  four correspondingly optimized  "stem  mixture  encoder-
decoders" that together provide our decomposition of the mixture input into its stem constituents. On a more technical note, in both training phases and deployment, the data is processed chunk-wise, with a size of about 9 seconds.

For a clear overview of the content of this section, refer to figure \ref{fig:fig1}. All conducted experiments that will be defined in the next section were computed on  two Tesla GPUs with 16Gb each. The length of each input sequence is equal to 393216 data points as used by \textit{Jukebox}. The batch size is equal to 4.

\begin{equation}
\label{eq:2}
\text{SDR}_{stem} = 10\;\text{log}_{10} \frac{\sum_n \Vert s(n)\Vert^2+\epsilon}{\sum_n \Vert s(n) - \hat{s}(n) \Vert^2+\epsilon}
\end{equation}

To benchmark the conducted experiments, signal-to-distortion ratio (SDR) metric is used, which is a common metric in other SOTA papers\cite{DBLP:journals/corr/abs-1909-01174}\cite{Stoeter2019}\cite{Hennequin2020}\cite{sawata2021all}\cite{stoller2018waveunet}. 
It is computed by equation (\ref{eq:2}), as stated in \cite{musicDemixing}, where $s(n)$ is the values of the ground truth and $\hat{s}(n)$ depicts the values of the prediction. A small value $\epsilon=10^{-7}$ is added to avoid division by zero. 'Total' SDR is the mean SDR for all stems.

\section{Experiments and Results}

\begin{table}[htb]
\centering
\resizebox{.9\hsize}{!}{

\begin{tabular}{ |c||c|c|c|c|c|c| }
 \hline
 \multicolumn{6}{|c|}{SDR Values} \\
 \hline
 Method& Drum & Bass & Other & Vocal & Total\\
 \hline
 Our Approach (i)   & 4.925 & 4.073 & 2.695 & 5.060 & 4.188 \\
 Our Approach trained from scratch (ii) & -0.002 & -0.087 & -0.026 & 0.00 & -0.028 \\
 Our Approach without finetuning(iii) & 1.06 & -1.072 & 0.79 & 0.33 & 0.279 \\

 \hline
\end{tabular}
}
\caption{Comparison of SDR values per stem and in total for three different versions of our approach.}
\label{tab:sdr_tab_our_approach}
\end{table}

\begin{table}[htb]
\centering
\resizebox{.7\hsize}{!}{

\begin{tabular}{ |c||c|c|c|c|c|c| }
 \hline
 \multicolumn{6}{|c|}{SDR Values} \\
 \hline
 Method& Drum & Bass & Other & Vocal & Total\\
 \hline
 DEMUCS  & 6.509 & 6.470 & 4.018 & 6.496 & 5.873 \\
 Our Approach   & 4.925 & 4.073 & 2.695 & 5.060 & 4.188 \\
 Wave-U-Net & 4.22 & 3.21 & 2.25 & 3.25 & 3.23 \\
ScaledMixturePredictor  & 0.578 & 0.745 & 1.136 & 1.090 	 & 0.887 	 \\
 \hline
\end{tabular}
}
\caption{Comparison of SDR values per stem and in total.  Our approach outperforms both the \textit{ScaledMixturePredictor}, the basic baseline in the Music Demixing Challenge \cite{musicDemixing} and \textit{Wave-U-Net} \cite{stoller2018waveunet}, a classic approach of source seperation in the waveform domain while \textit{Demucs}~\cite{DBLP:journals/corr/abs-1909-01174} achieves current SOTA performance on the Dataset.}
\label{tab:sdr_tab}
\end{table}


The main key point of this paper consists of demonstrating that it is
possible to get decent audio quality by using transfer learning. For
this, we did three different experiments (i), (ii), and (iii) on the four audio stems.
In experiment (i) we trained each audio stem's stem encoder SE from
scratch without using any pretraining values. For the second experiment
(ii) we trained the SE's with initial weights chosen as the pre-trained
weights of Jukebox. Experiment (iii) is conducted to show the impact of fine-tuning the SE. We skip fine-tuning in training phase 1 and use the weights of Jukebox as is and train a ME like in the other phases.


For all these VQ-VAE, we pick the checkpoint 80K and train the corresponding mixture encoders ME in phase 2. For
these, and in both experiments (i) (ii) and  (iii), we initialized their weights
randomly. For the first experiment, we found out that all results
are bad, and no good audio quality is reached for the extracted
stems. The SDR values are equal to or near 0 for all four stems. For
the  second  experiment,  the  model converges  after 32 hours of
training in total on two Tesla GPU units with 16GB of VRAM each. (iii) shows that fine-tuning in phase 1 is important.



\begin{figure}[htb]
\centering
 \includegraphics[scale=0.7]{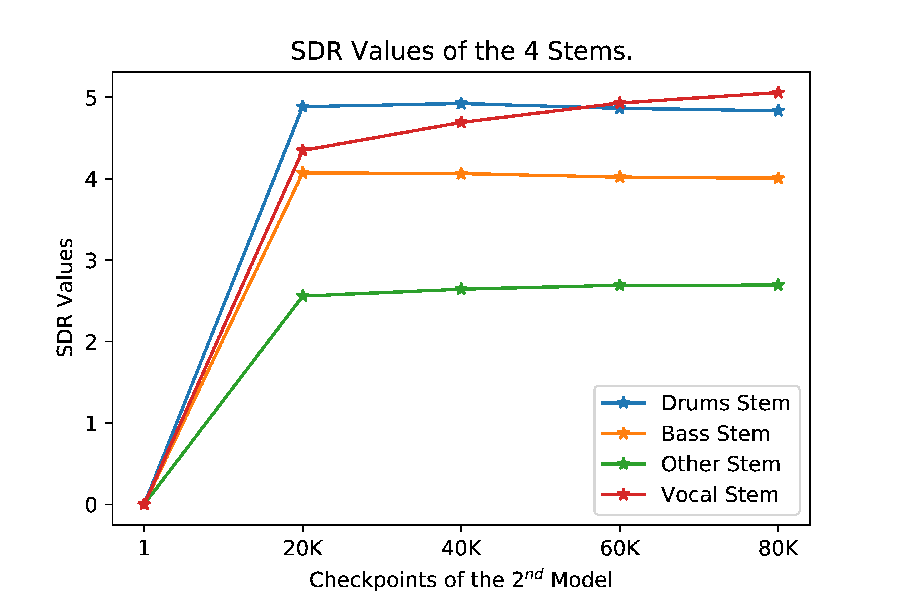}
 \caption{SDR results of the 4 audio signal stems for the second experiment.}
 \label{fig:res_1}
\end{figure}

Figure~\ref{fig:res_1} demonstrates decent SDR values for networks trained with  pre-trained weights in comparison to others trained with randomly initialized weights from scratch. It can also be deduced that in order to get fairly good SDR values, it is enough to train until early checkpoint values, such as 20K. Then, the checkpoint 20K is reached after 16 hours for each of the two models on two Tesla GPUs. Table \ref{tab:sdr_tab} gives a comparison of different approaches for audio signal separation. Here, our approach achieves comparable results when benchmarked against other state-of-the-art networks.

In terms of deployment, we need 0.73 seconds of CPU processing time for an audio chunk of 8.91 seconds per stem, which translates into an RTF (Real-Time Factor) of 0.082. That seems quick, but the current implementation of our model takes an audio chunk of 8.91 seconds. As with most approaches to audio source separation, direct online processing is not possible. However, a single code vector of the VQ-VAE represents a sliding window of an input vector of 49152 timesteps, which corresponds to about 1s of audio with a sampling rate of 44.1k. So, we think that an adapted architecture could be more flexible.


\section{Conclusion}



We demonstrate how transfer learning can be used for a problem of audio signal processing, in particular for the separation of an audio signal from a single mixed audio channel into four different stems: 'drums', 'bass', 'vocals', and 'other'. We show that it is possible to be successful with a small data set and relatively short training time on just two GPUs by fine-tuning pre-trained weights of \textit{Jukebox}~\cite{dhariwal2020jukebox}. Similar results could not be achieved in a comparable timeframe when training from scratch or without fine-tuning, showing the potential for reduced training times and improving results through the use of transfer learning in the audio domain.

\bibliographystyle{splncs04}
\bibliography{sources.bib}

\end{document}